\newcommand{\dif}{\mathrm{d}}
\newcommand{\I}{\mathbf{i}}
\newcommand{\J}{\mathbf{j}}

\newcommand{\m}{\mathbf{m}}

\newcommand{\R}{\mathbf{r}}

\newcommand{\nI}{n_{\I}}
\newcommand{\nJ}{n_{\J}}
\newcommand{\sI}{s_{\I}}
\newcommand{\sJ}{s_{\J}}
\newcommand{\qI}{q_{\I}}
\newcommand{\qJ}{q_{\J}}
\newcommand{\Hn}{{\mathcal H}_{\mathrm{N}}}
\newcommand{\HC}{{\mathcal H}_{\mathrm{C}}}
\newcommand{\betan}{\beta_{\mathrm{n}}}

\newcommand{\betaC}{\beta_{\mathrm{C}}}
\newcommand{\En}{E_{\mathrm{n}}}
\newcommand{\EC}{E_{\mathrm{C}}}
\newcommand{\HnC}{{\mathcal H}_{\mathrm{N+C}}}
\newcommand{\HFe}{{\mathcal H}_{\mathrm{Fe}}}

\newcommand{\M}{{\mathcal M}}

\newcommand{\sumNIJ}{\sum^V_{\I,\J}\hspace{-2pt}' }
\newcommand{\SUMNIdifJ}{\sum^V_{\I\neq\J}}
\newcommand{\SUMNIJ}{\sum^V_{\I , \J}}

\documentclass[prl,aps,twocolumn,psfig,preprintnumbers,amsmath,amssymb,showpacs]{revtex4}
\usepackage{graphicx}\usepackage{epsfig}
\begin{document}
\title{
    Ising analogue to compact-star matter
}
%
%
\author{P. Napolitani$^{1,2}$}
\author{Ph. Chomaz$^{1}$}
\author{F. Gulminelli$^{2}$\footnote{member of the Institut Universitaire de France}}
\author{K.H.O. Hasnaoui$^{1}$}
\affiliation{$^{1}$~GANIL (DSM-CEA/IN2P3-CNRS), Blvd. H. Becquerel, F-14076 Caen c\'edex, France}
\affiliation{$^{2}$~LPC (IN2P3-CNRS/Ensicaen et Universit\'e), F-14076 Caen c\'edex, France}
%
%
\begin{abstract}
	By constructing an Ising analogue of compact-star matter at 
sub-saturation density we explored the effect of Coulomb frustration 
on the nuclear liquid-gas phase transition.
	Our conclusions is twofold.
	First, the range of temperatures where inhomogeneous phases
form expands with increasing Coulomb-field strength.
	Second, within the approximation of uniform electron 
distribution, the limiting point upon which the phase-coexistence 
region ends does not exhibit any critical behaviour. 
	Possible astrophysics consequences and thermodynamical 
connections are discussed.
\end{abstract}
%
%
%
%
 \maketitle
%

	In absence of Coulombic interactions, the aspect of warm
nuclear matter below its saturation density is a dense phase
immersed in a low-density gas. 
	The Coulomb repulsion strongly affects this liquid-gas 
phase transition by forbidding the condensation in macroscopic 
drops.
    This phenomenon, originally described in condensed-matter
studies with the name of ``frustration'', denotes all systems
where no states exist where all interaction energies are 
simultaneously minimised: they range from magnets on specific 
lattices to liquid crystals, from spin glasses to protein folding.
%

	Nuclear matter in the density range of the liquid-gas 
phase transition can be produced either in the expansion and 
disassembling of nuclei involved in violent ion collisions, or in 
the core of supernovae explosions, as well as in inner neutron-star 
crusts.
	It is known since the seventies~\cite{Negele1973} that the 
clusterised solid configurations in the inner crust of a neutron 
star give rise to complex phases~\cite{pasta}, which are often 
quoted as pasta phases because of their suggestive topologies.  
	Recent molecular-dynamics simulations show that these
structures may survive also at finite temperature~\cite{mol_din}. 
	The fluctuations connected to pasta phases are expected to 
enhance matter opacity to neutrino scattering with important 
consequences on the supernova explosion and cooling 
dynamics~\cite{mol_din,Margueron2004}. 
	Such a coherent neutrino-matter scattering is not only 
expected at low temperature, but even more in the possible 
occurrence of a critical point in the post-bounce supernova 
explosion, with the associated phenomenon of critical 
opalescence~\cite{Margueron2004,wata_rev}.
	At variance with usual-matter properties, the expected 
increase in the static form factor was not observed in 
molecular-dynamics simulation of stellar matter at finite 
temperature~\cite{Horowitz2004}. 
	This might be an effect of the Coulomb interaction which 
also acts in finite nuclei. 
	However, it should be observed that, differently from 
nuclei, star matter includes also electrons with the role of 
neutralising the net charge over macroscopic portions.
%

     The motivation of the present work is  to provide general 
insights about the phase-transition phenomenology of neutral 
systems with Coulombic interactions.
	We construct an Ising analogue of compact-star matter by 
adding long-range interactions and a background of electrons to a 
Lattice-Gas model.
    Such an approach allows to take into account charge-density
fluctuations by exact calculations, and to compare stellar matter 
with respect to other physical systems subjected to frustration, 
like hot atomic nuclei\cite{Bonche1985,Gulminelli2003} and 
Coulomb-frustrated ferromagnets\cite{Grousson}.
%

	Let us consider a neutral system composed of charged 
particles occupying a cubic lattice of $V$ cells with occupation 
numbers $\nI=0,1$ immersed in a uniform background of negative 
charge representing the incompressible degenerated gas of 
electrons~\cite{Maruyama2005}.
 	This leads to a charge $\qI = \nI - \bar{n} $ for each site 
$i$, where $\bar{n} =\sum^{V}_{\J}\nJ/V$ comes from the uniform
background.
	The schematic Hamiltonian $\HnC=\Hn+\HC$ reads
\begin{equation}
	\Hn = \frac{\epsilon}{2} \sumNIJ \nI\nJ		
	,\quad
	\HC = \frac{\lambda\epsilon}{2}
	      \SUMNIdifJ \frac{\qI\qJ}{r_{\I\J}}		
	,
\label{eq:Hamiltonian}
\end{equation}
where $\sum'$ extends over closest neighbours. 
The isospin degree of freedom is not explicitly 
accounted, to allow a delocalization of the charge 
over the lattice\cite{dorso}.
The nuclear symmetry energy, here neglected, 
is known to change only quantitatively the phase 
diagram\cite{lattice_isospin}. 
	The short-range (nuclear-like $\Hn$) and long-range 
(Coulomb-like $\HC$) interactions are characterised by the coupling 
constants $\epsilon$ and $\lambda\epsilon$ respectively, so that
$\lambda$ measures the strength of frustration.
	To mimic stellar matter the numerical values are set to 
$\epsilon=-5.5 MeV$ and 
$\lambda\epsilon = \alpha\hbar c \rho_0^{1/3} x^2$, where 
$\rho_0=0.17 fm^{-1}$ is the nuclear saturation density, and 
$x=1/3$ is a typical proton fraction.
	$\HC$ can be rewritten as
$\HC = \frac{\lambda\epsilon}{2} \SUMNIJ \nI\nJ C_{\I\J}$, where
$C_{\I\J} = r^{-1}_{\I\J} - {\bar{r}}^{-1}$ if $\I\neq\J$ and 
$C_{\I\I} = - {\bar{r}}^{-1}$ with 
${\bar{r}}^{-1}=\frac{1}{V}\sum_{\mathbf{j}\neq\mathbf{0}}^{V} r_{\mathbf{0}\mathbf{j}}^{-1}$
and the distance $\R_{\I\J}$ is imposed to be the shortest 
between $\I$ and $\J$ in the periodic space.
To numerically accelerate thermodynamic convergence, the 
finite lattice is repeated in all three directions of space a large 
number $R$ of times, analogous to the Ewald summation technique.
    Each site $\I$ has then $R$ replicas of itself, each one 
displaced from $\I$ of a vector $\m L$, where $\m$ has integer 
components and $L = \sqrt[3]{V}$ is the cubic lattice length.
	This procedure is equivalent to a renormalisation of the 
long-range coupling $C_{\I\J}$~\cite{in-progress}.

	In ref.~\cite{Grousson} a complete thermodynamic study of 
frustration is dedicated to Ising ferromagnets, described by the 
Hamiltonian
\begin{equation}
	\HFe = \frac{\epsilon}{2} \sumNIJ \sI\sJ +
	\frac{\lambda\epsilon}{2} \SUMNIdifJ\frac{\sI\sJ}{r_{\I\J}}
	,
\label{eq:Fe}
\end{equation}
where $\sI=\nI-1/2$ is a spin variable. 
	The two models are related by:  
\begin{equation}
   \HFe = \HnC + \frac{\lambda\epsilon}{2{\bar{r}}} \M^2
        + \mu_{c}\M - \frac{3\epsilon}{4}V
	,
\label{eq:isomorphism}
\end{equation}
where $\M=\sum_{\I}^{V}\sI$ is the magnetisation.  
	We deduce that, contrarily to the standard form of 
closest-neighbour interactions, the charged Lattice-Gas model, 
eq.(\ref{eq:Hamiltonian}), in the grand canonical ensemble
cannot be mapped into an Ising-ferromagnet model, 
eq.(\ref{eq:Fe}), in an external magnetic field because of the term 
in $\M^2$ in eq.(\ref{eq:isomorphism}).
	Since this term scales with $V^2$, a thermodynamic limit 
exists for $\HFe$ only for vanishing magnetisation conversely to 
$\HnC$~\cite{Chomaz2005}.
Since $\M$, or the particle 
density $\rho=\M/V+1/2$, is an order parameter of the Ising model, 
to impose $\M=0$ as a strict constraint can substantially modify the 
thermodynamics of the system~\cite{Carmona2003}.
	Moreover, the presence of the non-linear term $\M^2$
directly affects the curvature of the order parameter 
distribution and so modifies the phase properties.
%
 
	Clarified these distinctions, henceforth we analyse the 
thermodynamics of the $\HnC$ Hamiltonian.
	
%
%
\begin{figure}[b!]
\begin{center}
\includegraphics[angle=0, width=1\columnwidth]{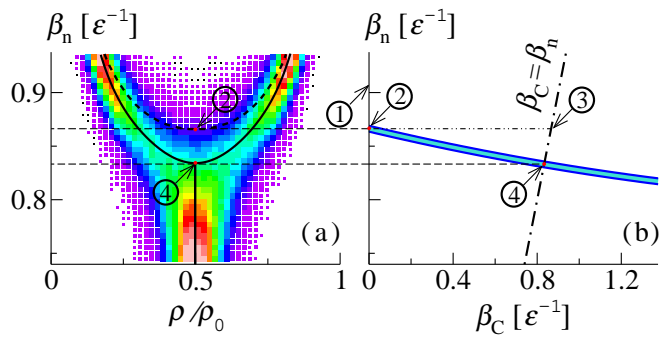}
\end{center}
\caption{\label{fig:fig1}
	(Color online)
	Metropolis calculations ($\mu=\mu_c$, $L=10$):
		\textbf{(A)}
	The logarithmic cluster plot gives the density 
	distributions for different values of $\betan$ and for 
	$\betaC=0.83\epsilon$, which is the inverse limiting 
	temperature computed for $\betaC=\betan$.
	The solid line indicates the ridge of the distribution,
	the dashed line  the coexistence region of an Ising-like 
	system ($\betaC=0$). 
	The points (2) and (4) indicate the limiting temperatures 
	for $\betaC=0$ and $\betaC=\betan$.
		\textbf{(B)}
	Phase diagram in the multicanonical ensemble giving the 
	inverse limiting temperature $\betan$ as a function of
	$\betaC$. The points (1) and (3) are used in 
	Fig.~\ref{fig:fig2}.
}
\end{figure}
%
%
%

	The multi-canonical ensemble \cite{Gulminelli2003} is a 
specifically adapted statistical tool to deal with frustrated 
systems.
	The Hamiltonian components $\Hn,\HC$ are treated as two 
independent observables associated to two Lagrange multipliers 
$\betan,\betaC$, respectively. 
	The multi-canonical partition sum reads
\begin{eqnarray}
	&&Z_{\betan,\betaC}(N,V)
	\\ &&\qquad = 
	\int W(\En,\EC,N,V) e^{-\betan\En-\betaC\EC} \dif\En\dif\EC   
	\notag , 
	\label{cano}
\end{eqnarray}
where $W(\En,\EC,N,V)$ is the density of states with nuclear energy
$\En$, Coulomb energy $\EC$, and number of particles $N$.
	A generalized grand potential is defined by
\begin{equation}
	Z^G_{\betan,\betaC,\alpha}(V)
	= \int Z_{\betan,\betaC}(N,V) e^{\betan \mu N}\dif N
	.
	\label{macro}
\end{equation}
	When $\betaC = \betan$ the ensemble coincides with the
conventional grandcanonical form.
	When $\betaC = 0$ the system reduces to the standard Ising 
model. 
	Therefore, the multi-(grand)canonical ensemble allows to 
construct one single phase diagram both for neutral and charged 
matter by exploring the space $(\betan,\betaC)$.
	It should be noticed that varying independently $\betan$ 
and $\betaC$ is equivalent to changing the effective charge
$q^2_{\mathrm{eff}} = \lambda\betaC/\betan$.
%

	The canonical phase diagram in the $(\betan,\betaC)$ space
can be accessed from the topological properties of the particle
density distribution in the corresponding grand-canonical 
ensemble eq.(\ref{macro})~\cite{Chomaz2002}. 
	For a given value of $\betaC$, Fig.~\ref{fig:fig1}A 
describes, as a function of the inverse temperature $\betan$, the 
grand-canonical density distribution calculated at the critical 
chemical potential $\mu=\mu_c$.
	This distribution is sampled from eq.(\ref{macro}) with a 
standard Metropolis technique~\cite{Chomaz2002}.
	At high $\betan$, the density distribution shows two peaks 
of the same height, which are associated to two coexisting 
phases~\cite{Chomaz2002}.
	These two phases join at
the limiting temperature (point (4)) 
above which the distribution stops to be 
bimodal.
	From a series of similar calculations performed for 
different values of $\betaC$, we could extract the evolution of the
inverse limiting temperature as a function of $\betaC$ as plotted 
in Fig.~\ref{fig:fig1}B.
	We observe an increase of the limiting temperature
$\betan^{-1}$ for increasing strength of the Coulomb field
(about 6\% for a proton fraction $x=1/3$).
%

	It is important to notice that many other physical systems 
subjected to Coulomb frustration exhibit the opposite behaviour.
This is notably the case of frustrated Ising ferromagnets, 
as well as of finite atomic nuclei.
	In such cases the Coulomb repulsion is known to reduce the 
limiting temperature~\cite{Bonche1985,Lee2001,Grousson,Raduta2002,Gulminelli2003}.
	This reduction is also an usual expectation in the 
astrophysical context~\cite{pasta,Lattimer1985,Haensel2000}.
	However, a recent calculation of nuclear-pasta structure 
within the RMF model~\cite{Maruyama2005} indicated a widening of 
the density range connected to the mixed phase region when the 
Coulomb field is included under the constraint of charge neutrality 
over the Wigner-Seitz cells.
	Our results are consistent with this finding, and suggest 
that this effect is model independent and should persist at finite 
temperature\cite{Chomaz2005}.
%

%
%
\begin{figure}[b!]
\begin{center}
\includegraphics[angle=0, width=0.80\columnwidth]{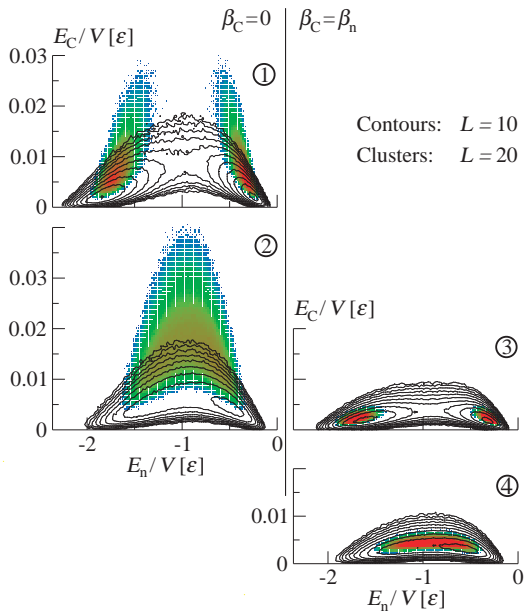}
\end{center}
\caption{\label{fig:fig2}
	(Color online)
	Probability distribution for the two energy components 
	$\En,\EC$ for the four points indicated in 
	Fig.~\ref{fig:fig1}B.
	Contour plots refer to $L=10$; logarithmic cluster plots 
	refer to $L=20$.
	Points (1) and (3) belong to the coexistence region and
	manifest bimodal patterns.
	Both point (2) and (4) are very close to the limiting temperature
	but the critical scaling effects exhibited at point (2)
	are absent at point (4).
}
\end{figure}
	The analysis of the event distribution 
elucidates the physical origin of 
the increase of the limiting temperature.
The left side of Fig.~\ref{fig:fig2}
shows the transition from subcritical to critical partitions
for an uncharged system ($\betaC = 0$).
When the system is uncharged, the observable $\EC$ defined 
in eq.(\ref{eq:Hamiltonian}) does not represent a physical 
Coulomb energy, and measures the compactness of the system. 
 Its contribution does not influence the partition 
probability since $\betaC = 0$, and 
compact	clusterised partitions with high $\EC$
can be explored.
%

When the system is charged (right side of Fig.~\ref{fig:fig2}), 
$\EC$ enters in the partition distribution eq.(\ref{cano}) as
the physical Coulomb energy. This energy is maximal in fragmented 
configurations at $\rho/\rho_0\approx 0.5$, and minimal 
in pure-phase events, where the largely 
uniform proton-charge distribution is almost exactly compensated by 
the electron background. The pure-phase events are then increasingly favored
with increasing charge. Thus, bimodal distributions are still found 
in the charged system (panel 3) at temperatures at which critical events
dominate the uncharged system (panel 2).
This is entirely due to the screening effect of the electrons, 
while in finite nuclei the highest Coulomb energy is associated to 
the homogeneous liquid-like partitions, leading to the opposite effect.
%

	The question now arises as to whether the limiting points
of the frustrated system are second-order critical points like in 
the uncharged system, or in the Coulombic RPM model describing 
phase separation in electrolytes \cite{kim}.
	Criticality arises from the divergence of
the correlation length $\xi$.  
	It describes the exponential decay of the correlation 
function $\sigma(\mathbf{r}_{\mathbf{i},\mathbf{j}}) = \langle\delta\nI \delta\nJ \rangle =
\langle \nI\nJ \rangle - \langle \nI \rangle \langle \nJ \rangle$
according to the expression  
$\sigma(r) \propto   e^{-r/\xi} \cdot r^{-(D-2+\eta)}$,
where $D$ is the space dimension and $\eta$ is a critical exponent.
	Using the properties of the $C_{\I\J}$ matrix, $\sigma(r)$
can be easily related to the mean Coulomb-energy density by 
\begin{equation}
	\Big\langle\frac{\EC}{V}\Big\rangle =
	\frac{\sigma(0)}{\bar{r}} + 
	\frac{\lambda\epsilon}{2} \sum_{\mathbf{j} \neq \mathbf{0}}^{V}
\frac{\sigma(\mathbf{r}_{\mathbf{0}\mathbf{j}})}{r_{\mathbf{0}\mathbf{j}}}
        .
\label{eq:energydens}
\end{equation}
	Eq.(\ref{eq:energydens}) indicates that a diverging 
correlation length at the critical point manifests by a divergent 
Coulomb-energy density.
When $\betaC \neq 0 $ 
	the only solution for the system to avoid such a 
singularity is to suppress the second-order critical character of 
the limiting point~\cite{Chomaz2005}. 
	In this case, the correlation length keeps a finite value
at the limiting temperature and the thermodynamic limit is 
fulfilled by a constant value of $\langle\EC/V\rangle$. 
	Such a behaviour is tested in Fig.~\ref{fig:fig2}, where 
the same calculation is repeated for different lattice sizes:
the average Coulomb energy is seen to increase with the lattice 
size in the Ising system,
while Coulomb-energy fluctuations appear quenched in 
the frustrated system. 
It is interesting to note that this argument does not apply in other
neutral Coulombic systems \cite{kim}, where density fluctuations
do not necessarily imply charge fluctuations, and can therefore diverge
keeping a finite Coulomb energy. For the same to be true in the proto-neutron
star, the electron field should be strongly polarized at the limiting 
temperature, leading to a 
complete charge screening of the dishomogeneous pasta structures. 
Due to the high incompressibility of the degenerate electron gas, this is likely to be unphysical\cite{Maruyama2005}.
%

	The quenching of criticality can be formally verified in terms of 
critical exponents.
	If the asymptotic value of the limiting temperature 
$T_{\mathrm{lim}}=\lim_{L\rightarrow\infty} \dot{T}_{\mathrm{lim}}(L)$ 
corresponds to a critical point, finite-size scaling insures that 
$\dot{T}_{\mathrm{lim}}(L)$ should evolve as 
$\dot{T}_{\mathrm{lim}}(L) - T_{\mathrm{lim}} \propto L^{-1/\nu}$~\cite{scaling}.
%
%
\begin{figure}[bp!]
\begin{center}
\includegraphics[angle=0, width=0.9\columnwidth]{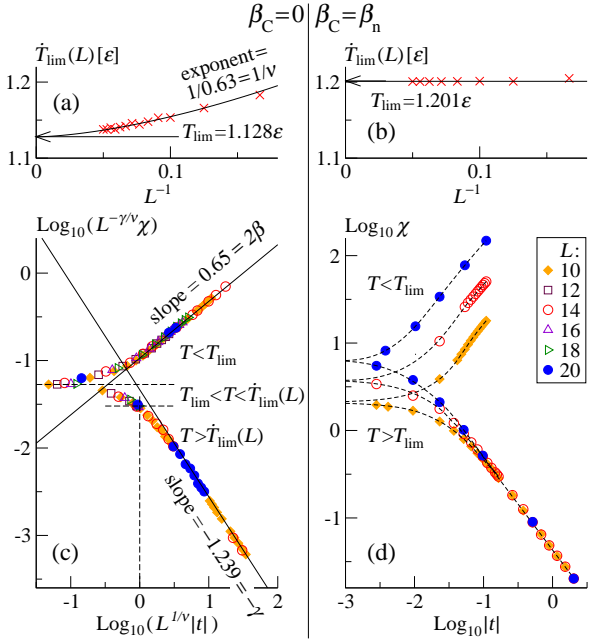}
\end{center}
\caption{\label{fig:fig3}
	(Color online).
	Panels A,B. Evolution of $T_{lim}$ as a function of the 
	linear size $L$, and extrapolation towards the 
	thermodynamic limit.
	Panels C,D. Study of the finite-size scaling 
	eq.(\ref{eq:hyperscaling}).
	All calculations are presented for the uncharged 
	($\betaC=0$) and the charged system ($\betaC=\betan$) with 
	proton fraction $x=1/3$.     
}
\end{figure}
	Fig.~\ref{fig:fig3} 
	illustrates that, 
while the scaling is respected with the Ising value of $\nu$ 
when $\betaC=0$, $\nu$ should be infinitely large to describe
the frustrated system.
	Since $\nu$ rules the divergence of the correlation length,
$\xi\propto t^{-\nu}$ with 
$t = T/T_{\mathrm{lim}}-1$, this is a first indication of a finite 
correlation length for the charged system. 
	We then test the finite-size scaling theory~\cite{scaling}
on the quantity $\chi = \sum^{N}_{\I\J} \sigma_{\I\J} /T$.
	This form of $\chi$ represents the susceptibility only for 
$T\geq\dot{T}_{\mathrm{lim}}(L)$.
	In this case, it should scale with the critical exponent 
$\gamma$ as $L^{-\gamma/\nu}\chi = f(L^{-1/\nu} t)$ if the 
limiting point represented a critical point; 
the function $f(L^{-1/\nu} t)$ should be constant when $\xi\sim L$ 
and $f(L^{-1/\nu} t) \propto (L^{1/\nu} t)^{-\gamma}$ when 
$\xi\ll L$.
	When $T<\dot{T}_{\mathrm{lim}}(L)$, $\chi$ contains also
jumps between the low-density, $\langle n \rangle_{\mathrm{G}}$,
and the high-density, $\langle n \rangle_{\mathrm{L}}$,
solutions and should therefore scale with the critical 
exponent $\beta$ defined from the scaling of the order parameter 
$\langle n \rangle_{\mathrm{L}}-\langle n \rangle_{\mathrm{G}} 
\propto t^\beta$ as $\chi \propto L^D t^{2\beta}$.
	By introducing the hyperscaling relation
$D = (\gamma+2\beta)/\nu$, all scaling laws condense in the form
\begin{equation}
	L^{-\gamma/\nu}\chi \propto
	\left\{ \begin{array}{ll}  
		\mathrm{constant}	& \mathrm{as}\; \xi\sim L\\
		(L^{1/\nu} t)^{-\gamma}	& \mathrm{as}\; \xi\ll L,\; 
			T\geq\dot{T}_{\mathrm{lim}}(L)\\
		(L^{1/\nu} t)^{2\beta}	& \mathrm{as}\; \xi\ll L,\; 
			T<\dot{T}_{\mathrm{lim}}(L)
	\end{array}\right.
\label{eq:hyperscaling}
\end{equation}
	Fig.~\ref{fig:fig3}C illustrates the perfect consistency of
the $\betaC=0$
system with Ising critical exponents.
	 Conversely, for the charged system, no combination of
$\beta$, $\gamma$ and $\nu$ can be found to let all calculated
points collapse on the same law.
	In particular, for $\gamma/\nu=0$ and $\nu\to\infty$, only
the points for $T\gg\dot{T}_{\mathrm{lim}}(L)$ collapse on one 
line, while those for $T\ll\dot{T}_{\mathrm{lim}}(L)$ disperse as
shown in Fig.~\ref{fig:fig3}D.
	This demonstrates that the effect of the Coulomb field is 
not a simple increase of the $\nu$ exponent, but a complete 
quenching of criticality for the frustrated system.
	The loss of critical behaviour has been already observed in 
Ising models for the Ising frustrated ferromagnet, 
eq.(\ref{eq:Fe}), where the coexistence  region was seen to end at 
a first-order point~\cite{Grousson}.

%

	Our conclusion is twofold.
First, we observed that,
in presence of a uniform electron background, the Coulomb field 
originating from charge-density fluctuations 
increases 
the limiting temperature for phase coexistence, at variance with the 
strong decrease obtained in the absence of electrons or in 
mean-field calculations~\cite{Lattimer1985,Haensel2000}.
	Therefore, the mixed-phase phenomenology may be relevant 
for the proto-neutron-star structure 
up to slightly above the critical temperature of normal nuclear matter 
($T_c\approx 15$ MeV for symmetric matter), 
in a wider temperature-range 
than usually expected~\cite{pasta}.
	Second, the Coulomb field suppresses the critical character 
of the limiting temperature.
  	For this reason we expect warm stellar matter to show 
small opacity to neutrino scattering in agreement with 
the findings of ref.~\cite{Horowitz2005}.
%
%

%
%
%
%
\end{document}